\documentclass[conference]{IEEEtran}

\IEEEoverridecommandlockouts
\usepackage{cite}
\usepackage{enumitem}

\usepackage[font=small,labelsep=period]{caption}

\usepackage{algorithm}
\usepackage{algorithmic}

\usepackage{booktabs} 
\usepackage{multirow} 

\usepackage{amsmath,amssymb,amsfonts}
\usepackage{algorithmic}
\usepackage{url}
\usepackage{gensymb} 

\usepackage{graphicx}
\usepackage{textcomp}
\usepackage{xcolor}
\def\BibTeX{{\rm B\kern-.05em{\sc i\kern-.025em b}\kern-.08em
    T\kern-.1667em\lower.7ex\hbox{E}\kern-.125emX}}
\begin{document}

\title{Mitigating Image Captioning Hallucinations in Vision-Language Models 
}
\author{\IEEEauthorblockN{Anonymous ICME Submission}}


\author{
\IEEEauthorblockN{
Fei Zhao\IEEEauthorrefmark{1},
Chengcui Zhang\IEEEauthorrefmark{1},
Runlin Zhang\IEEEauthorrefmark{2},
Tianyang Wang\IEEEauthorrefmark{1},
Xi Li\IEEEauthorrefmark{1}
}
\IEEEauthorblockA{\IEEEauthorrefmark{1}Department of Computer Science, The University of Alabama at Birmingham, Birmingham, USA\\
\{larry5, czhang02, tw2, xiliuab\}@uab.edu}
\IEEEauthorblockA{\IEEEauthorrefmark{2}Department of Computer Science, University of Waterloo, Waterloo, Canada\\
r496zhan@uwaterloo.ca}
}

\maketitle

\maketitle

\begin{abstract}
Hallucinations in vision-language models (VLMs) hinder reliability and real-world applicability, usually stemming from distribution shifts between pretraining data and test samples. Existing solutions, such as retraining or fine-tuning on additional data, demand significant computational resources and labor-intensive data collection, while ensemble-based methods incur additional costs by introducing auxiliary VLMs. To address these challenges, we propose a novel test-time adaptation framework using reinforcement learning to mitigate hallucinations during inference without retraining or any auxiliary VLMs. By updating only the learnable parameters in the layer normalization of the language model (approximately 0.003\% of the model parameters), our method reduces distribution shifts between test samples and pretraining samples. A CLIP-based hallucination evaluation model is proposed to provide dual rewards to VLMs. Experimental results demonstrate a 15.4\% and 17.3\% reduction in hallucination rates on LLaVA and InstructBLIP, respectively. Our approach outperforms state-of-the-art baselines with a 68.3\% improvement in hallucination mitigation, demonstrating its effectiveness.
\end{abstract}


\section{Introduction}
\label{sec:intro}
VLMs have become foundational for tasks such as image captioning and visual question answering (VQA), demonstrating remarkable capabilities in aligning textual and visual modalities \cite{zhao2024deep}. However, these models often suffer from a critical issue: hallucinations, where the generated output deviates from the input image content. Such hallucinations undermine the reliability and applicability of VLMs in real-world scenarios, particularly in high-stake domains including autonomous systems, medical image analysis, and surveillance, where factual accuracy is paramount \cite{liu2024survey}.

The root cause of hallucinations often lies in distribution shifts between pretraining data and real-world test samples. Pretraining on noisy datasets introduces biases and reliance on unimodal priors, leading models to generate hallucinated content when presented with unseen or domain-specific data. Addressing this challenge requires methods that adapt models to new data distributions while ensuring computational efficiency. Existing methods to mitigate hallucinations, including retraining or fine-tuning, ensemble approaches, and logit manipulation, face limitations in scalability, computational efficiency, and generalizability (see Section~\ref{sec:related_work}). 

In this work, we propose a novel test-time adaptation (TTA) framework using reinforcement learning (RL) to dynamically mitigate hallucinations during inference. In our approach, the VLM itself acts as the policy model in an RL framework, allowing it to iteratively refine its output captions based on feedback from a hallucination evaluation model. RL is well-suited to this task as it enables the model to optimize its decisions (caption generation) based on feedback signals (rewards) rather than requiring extensive retraining. The policy model learns to adapt to each test sample individually, reducing distribution shifts on a per-sample basis.

The key contributions of this work are as follows:
\begin{enumerate}[label=\arabic*., left=0pt]
    \item \textbf{A Novel TTA Framework for Hallucination Mitigation:} 
    We introduce a lightweight and efficient TTA method using RL to reduce the data distribution shift between test samples and pretraining data. It \textbf{mitigates hallucinations in VLMs during inference}, without requiring retraining or auxiliary VLMs.

    \item \textbf{A Lightweight Hallucination Evaluation Model:} 
    We propose a CLIP-based hallucination evaluation model equipped with a learnable query prompt to extract knowledge from the frozen CLIP encoders. This model is capable of independently classifying image-text pairs as non-hallucinated or hallucinated. Furthermore, it serves as a reward model in RL, providing dual rewards: \textit{Semantic Alignment Score} (SAS) for image-text alignment and \textit{Non-Hallucination Probability} (NHP) scores for the likelihood of non-hallucination. While SAS measures semantic alignment, it does not ensure the caption is free from hallucinations. NHP complements SAS by assessing factual consistency, addressing hallucination. Together, these dual rewards guide the policy model to dynamically refine the generated content, achieving superior performance in mitigating hallucinations.

\item \textbf{A Parameter-Efficient RL Approach for VLMs:}  
Our approach updates only the learnable parameters in layer normalization \cite{xu2019understanding}, accounting for approximately 0.003\% of the model’s total parameters. This is significantly more efficient than modifying entire VLMs or cross-modal projection layers that have nearly 100 times more parameters. 


\end{enumerate}


  \begin{figure*}[ht]
    \centering

    \includegraphics[width=\textwidth]{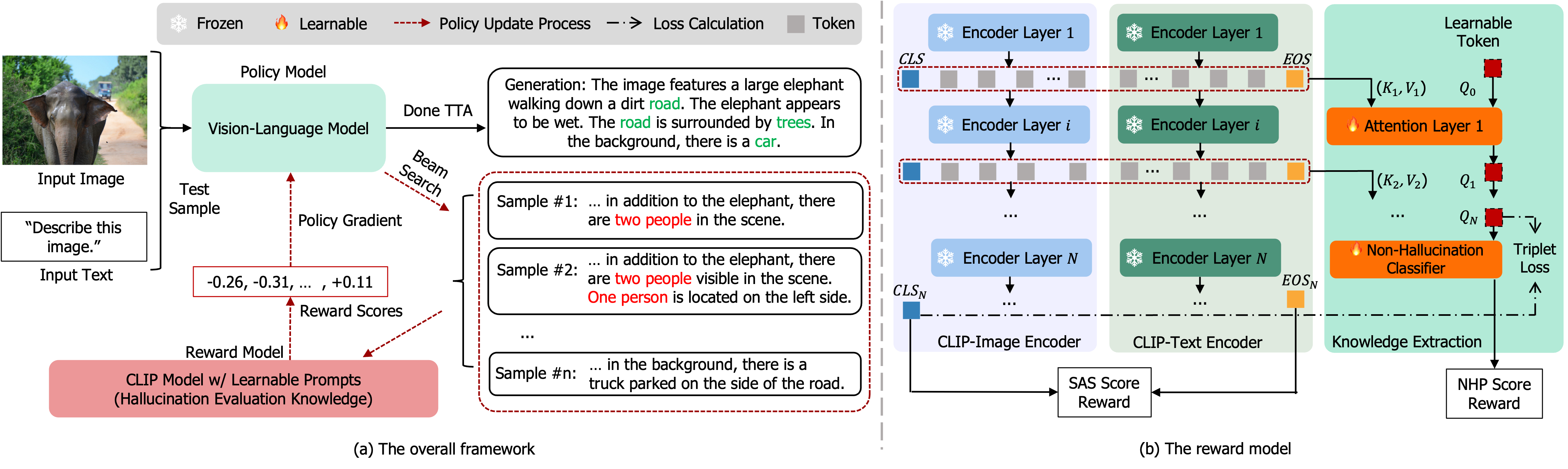}

    \caption{(a) The overall framework for mitigating object hallucinations in VLMs using TTA. The red texts indicate object hallucinations in the generated captions before TTA, while the green text highlights accurate descriptions appearing after TTA. (b) The architecture of the hallucination evaluation model, which leverages a frozen CLIP model with learnable query prompts, multi-stage cross-attention layers, and a binary classifier. The model facilitates training through triplet loss to differentiate between non-hallucinated and hallucinated captions.}

    ~\label{fig:overall}

  \end{figure*}

\textbf{Scope}: This paper focuses on mitigating object hallucinations in VLMs, specifically addressing captions with incorrect or non-existent objects. Attribute and relation hallucination mitigation is beyond the scope of our research for two reasons: first, there are too few benchmarks for evaluating these types of hallucinations, many of which rely on non-open-source APIs that lack accessible resources; second, our primary goal is to compare with the SOTA method VCD \cite{leng2024mitigating}, which primarily targets object hallucination mitigation.


The remainder of this paper is organized as follows: 
Section~\ref{sec:related_work} reviews related work on hallucination mitigation in VLMs. 
Section~\ref{sec:methodology} details our proposed methodology, including the TTA framework, the RL setup, and the customized CLIP-based hallucination evaluation model. 
Section~\ref{sec:datasets} describes the datasets, 
while Section~\ref{sec:experiments_results} presents experimental results and offers ablation studies to further highlight the contributions of individual components. 
Finally, Section~\ref{sec:conclusion} concludes the paper and discusses potential future directions.

\section{Related Work}
\label{sec:related_work}

VLMs have made significant progress in multimodal tasks, including image captioning and visual question answering (VQA). Proprietary state-of-the-art (SOTA) VLMs such as ChatGPT \cite{openai2023chatgpt}, Gemini \cite{deepmind2023gemini}, and Claude demonstrate impressive multimodal capabilities but are not open source, restricting their accessibility for research and customization. In contrast, open-source VLMs such as LLaVA \cite{liu2023llava} and InstructBLIP \cite{huang2023visual} offer competitive performance. Currently, VLMs are widely adopted for real-world applications. For instance, LLaVA has been effectively applied to extract financial information from check images \cite{zhao2024checkguard}. Despite these advancements, VLMs are prone to hallucinations, where the generated output deviates from visual input, particular under distribution shifts between pretraining and real-world test data.

Current approaches for mitigating hallucinations can be grouped into four categories: \textbf{(1) Visual input enhancement}, focusing on improving the quality of visual features fed into VLMs. For example, methods like LLaVA-Next \cite{liu2024llavanext} and InternVL \cite{chen2023internvl} adopt a multi-scale strategy: dividing the input image into smaller patches, resizing these patches to higher resolutions to emphasize local details, and downscaling the original image to capture global context. These processed images are then concatenated to create a richer visual representation. Similarly, Prismer \cite{liu2023prismer} incorporates auxiliary vision experts/submodels to further diversify visual inputs and enhance performance. While effective, these methods require complete retraining of the VLM, which is computationally expensive and impractical for refining already-deployed models. \textbf{(2) Fine-tuning VLMs}, typically on domain-specific datasets, has shown promise in mitigating hallucinations. For instance, the authors in \cite{chen2024detecting} fine-tuned LLaVA \cite{liu2023llava} for medical applications using newly collected domain-specific VQA data, significantly improving performance. Similarly, the authors in \cite{zhao2023test} apply RL to fine-tune VLMs such as CLIPCap \cite{mokady2021clipcap}, focusing on training the vision-language projector (\textbf{42 million parameters}). While effective, fine-tuning VLMs requires extensive computational resources and domain-specific data collection, making it resource-intensive and challenging to scale. \textbf{(3) Ensemble-based methods}, usually aggregating outputs from multiple VLMs or combining a weaker model with a stronger one. For instance, the authors in \cite{yu2024mitigating} propose a debate-style ensemble framework where two VLMs iteratively refine their outputs. Meanwhile, the authors in \cite{xiao2024detecting} leverage ChatGPT to provide detailed feedback for refining LLaVA outputs, reducing hallucination rates. While these techniques improve robustness, they are computationally costly. Furthermore, using a stronger model to enhance a weaker one often negates the practical benefits of deploying the less capable model. \textbf{(4) Logit manipulation methods}, modifying output distributions to mitigate hallucinations. The work \cite{zhao2024mitigating} employs an external vision encoder to provide soft prompts, mitigating hallucinations without additional training. However, without fine-tuning, the external encoder may produce visual features that are misaligned with the vision-language projector, potentially introducing noise and degrading performance. Similarly, Visual Contrastive Decoding (VCD) \cite{leng2024mitigating} reduces hallucinations by contrasting logits generated from original and noise-distorted inputs, assuming that matching outputs across the two inputs indicate hallucination. While this approach shows promise for single-token prediction tasks, it faces significant limitations in auto-regressive generation. The assumption that identical output tokens from those two inputs are hallucinated breaks down in sequential tasks, as the distorted input cannot guarantee that all its outputs are incorrect or hallucinated. For instance, common words such as ``this,'' ``that,'' ``is,'' or ``are'' might appear in both the original and distorted outputs, but they are often correct and contextually appropriate. Misclassifying these correct tokens as hallucinated can lead to unnecessary modifications, disrupting the generation process and introducing cascading errors that negatively affect subsequent tokens. These limitations make VCD less scalable and effective for auto-regressive applications.


\section{Methodology}
\label{sec:methodology}







As we mentioned in Section \ref{sec:intro}, we treat the VLM as a policy model that iteratively refines its output captions based on feedback from a hallucination evaluation model, which serves as the reward provider. This framework operates entirely during inference, eliminating the need for retraining or auxiliary models, and dynamically adapts the VLM to test samples.

As shown in Fig.~\ref{fig:overall}, the process begins with the policy model generating multiple candidate captions for a given test image and a prompt ``Describe this image'' using beam search \cite{meister2020best}. Each candidate is combined with the input text image and evaluated by the reward model, which generates dual reward scores: a CLIP-based score for image-text alignment and a logit score for non-hallucination likelihood. These scores are aggregated to form a final reward signal. The policy model is then updated using policy gradient loss to improve its captioning decisions. This iterative process continues for a predefined number of steps, culminating in the generation of a refined caption. The entire process is outlined in Algorithm~\ref{alg:tta}.

\begin{algorithm}[h]
\caption{TTA with RL for VLMs Hallucination Mitigation}
\label{alg:tta}
\begin{algorithmic}[1]
\REQUIRE Pretrained VLM, CLIP$_{\text{Prompts + Triplet}}$, Visual Input \(v\), Textual Input \(x\), Steps \(num\_steps\), Beam Size \(B\)
\ENSURE Refined Caption \(y^*\)
\STATE \textbf{Initialize:} Freeze VLM parameters except LayerNorm gamma. Load pretrained CLIP$_{\text{Prompts + Triplet}}$.
\FOR{\(step = 1 \to num\_steps\)}
    \STATE   Generate \(B\) candidate captions using beam search.
    \STATE   For each caption, compute:
    \begin{itemize}
        \item SAS score for image-text alignment.
        \item NHP score for non-hallucination likelihood.
    \end{itemize}
    \STATE   Aggregate and normalize reward scores. (see Section \ref{subsection:reward})
    \STATE   Update LayerNorm gamma using policy gradient optimization.
\ENDFOR
\STATE   Generate the refined caption \(y^*\) using the updated VLM.
\RETURN \(y^*\)
\end{algorithmic}
\end{algorithm}

\subsection{Reinforcement Learning for Vision-Language Models}
 
In our framework, the auto-regressive VLM acts as the \textit{policy model}, parameterized by \(\theta\), and the task of generating a caption for an input image is treated as a sequential decision-making problem. The key components are:

\begin{itemize}
    \item \textbf{State \(s\)}: The state represents the context available to the model at each step \(t\), which includes:
    \begin{itemize}
        \item \(v\): The visual input (e.g., images).
        \item \(x\): The textual input (e.g., ``Describe this image'').
        \item \(y_{<t}\): The sequence of tokens generated so far up to token generating step \(t\).
    \end{itemize}
    Thus, \(s_t = \{v, x, y_{<t}\}\), encapsulating all information available for generating the next token.

    \item \textbf{Action \(a_t\)}: The next token \(y_t\) is generated at step \(t\).

    \item \textbf{Policy \(\pi(a|s;\theta)\)}: The VLM defines a probabilistic policy over actions given the state. At each step \(t\), the probability of generating a token \(y_t\) is:
{\small    \[
    \pi(y_t | s_t; \theta) = {softmax}(f(v, x, y_{<t}; \theta)),
    \]}
    where \(f\) is the VLM’s output logits.

    \item \textbf{Reward \(r\)}: After generating the entire caption \(y = \{y_1, y_2, \dots, y_T\}\), a reward \(r(y, v)\) is calculated by the hallucination evaluation model (see Section \ref{subsection:reward}).
\end{itemize}

The goal of RL is to maximize the expected reward \(J(\theta)\), defined as:
{\small\[
J(\theta) = \mathbb{E}_{\pi_\theta} [r(y, v)],
\]}
where \(r(y, v)\) is the reward assigned to the caption \(y\). To optimize \(J(\theta)\), we use the \textit{policy gradient method} \cite{sutton1999policy}, which computes the gradient of the objective function with respect to the model parameters \(\theta\):
{\small\[
\nabla_\theta J(\theta) = \mathbb{E}_{\pi_\theta} \left[\nabla_\theta \log \pi(y|s; \theta) \cdot r(y, v)\right].
\]}
This equation adjusts the model parameters to increase the likelihood of generating captions that maximize the reward.

During TTA, only the \textit{gamma parameters in layer normalization} are updated, while the rest of the model is frozen. The updates are guided by the policy gradient loss, defined as:
{\small\[
\mathcal{L}_{\text{policy}} = - \mathbb{E}_{\pi_\theta} \left[\log \pi(y|s; \theta) \cdot r(y, v)\right].
\]}

This formulation converts the maximization problem into a minimization problem by introducing the negative sign, aligning with standard optimization procedures. The logarithm (\(\log\)) plays a key role in shaping the model's behavior:

\begin{itemize}
    \item \textbf{For positive rewards} (\(r(y, v) > 0\)): Minimizing the loss follows the \(-\log\) curve, which pushes the logit higher, increasing the probability of generating this output.
    \item \textbf{For negative rewards} (\(r(y, v) < 0\)): Minimizing the loss moves the logit along the \(\log\) curve to the left, reducing its probability.
\end{itemize}

This interplay between the reward and the \(-\log\) curve allows the model to reinforce high-reward outputs and suppress low-reward ones, creating a self-correcting RL loop that refines captions effectively.

\subsection{Reward Calculation}
\label{subsection:reward}

In our framework, the reward signal combines two components to guide the VLM toward generating accurate and semantically aligned captions. The first component is the \textit{Semantic Alignment Score} (SAS), which evaluates the cosine similarity between the visual input \({v}\) and the generated caption \({y}\) based on their embeddings from the hallucination evaluation model's frozen vision and text encoders, shown in Fig. \ref{fig:overall}. The second component is the \textit{Non-Hallucination Probability} (NHP), derived from the hallucination evaluation model's classification subnetwork that predicts the likelihood of the caption being non-hallucinated. The reward values are normalized by subtracting the mean across all candidate captions. The final reward \(r\) is computed as:
{\small\[
r(y, v) = \text{norm\_SAS} + \beta \cdot \text{norm\_NHP}
\]}where \(\beta\) is a weighting factor, empirically set to 1. This dual-reward system ensures that the VLM generates captions that are more semantically aligned with the input image with less hallucination, making it a crucial component of the RL framework.

\subsection{Hallucination Evaluation Model}
\label{subsection:non_hallucination}

The hallucination evaluation model builds on CLIP \cite{radford2021learning}, a widely used standard for image-text alignment \textbf{with strong zero-shot generalization}. We introduce a single learnable query token \({Q}\) atop the frozen CLIP architecture. This token is updated iteratively through \(N\)-stage cross-attention \cite{vaswani2017attention} layers to extract alignment and semantic consistency features between image and text inputs. The final updated query token features (\({Q}_{{N}}\)) serve as input to a classifier, composed of stacked fully connected layers, which predicts whether an image-text pair is non-hallucinated (positive) or hallucinated (negative). The classifier’s output, a logit score referred to as the NHP score, serves as a crucial reward in the RL framework.

\textbf{Learnable Query Tokens:} The single learnable query token \({Q}\) interacts with image and text features through cross-attention. At each stage \(i\), the tokens are updated as:
{\small
\[
{Q}_i = \text{CrossAttention}({Q}_{i-1}, {K}_{i}, {V}_{i}), \quad i \in \{1, 2, \dots, N\},
\]}
where \({Q}_{i-1}\) is the query token from the previous stage, and \({K}_{i}, {V}_{i}\) represent the keys and values derived from the concatenated image (\(v\)) and text (\(y\)) features at stage \(i\). The CLS token (classification token representing the image's global features) from the image encoder and the EOS token (End-of-Sequence token) from the text encoder are used to compute the SAS score via their cosine similarity, providing a quantitative measure of alignment between the image and the caption.

\textbf{Triplet Loss:} To guide the learnable query token \({Q}\) in extracting meaningful features from frozen vision and textual encoders, we incorporate a triplet loss~\cite{schroff2015facenet} with a margin \(\alpha\). The triplet comprises the image as the anchor, a ground truth (positive) caption, and a generated (negative) caption that contains hallucinations (details in Section~\ref{sec:datasets}). Let \({CLS}_{N}\) denote the final CLS token, and \({Q}_{N}^{\text{pos}}\) and \({Q}_{N}^{\text{neg}}\) be the updated query tokens for the positive and negative captions, respectively. The loss is defined as:
{\small
\[
\mathcal{L}_{\text{triplet}} = \max\left(0, \cos\left({CLS}_{N}, {Q}_{N}^{\text{neg}}\right) - \cos\left({CLS}_{N}, {Q}_{N}^{\text{pos}}\right) + \alpha\right)
\]
}
where \(\cos\) represents cosine similarity. The triplet loss ensures that the query token extracts discriminative features for the subsequent binary classification.

\section{Dataset and Experiments}
\label{sec:datasets}

This work focuses on mitigating object hallucinations in image captioning tasks. To achieve this, we utilize the {AMBER} \cite{wang2023llm} dataset for evaluation and a curated subset of the {PixelProse} \cite{singla2024pixels } dataset for training our customized CLIP-based hallucination evaluation model.

The {AMBER} dataset contains 1,004 test samples designed specifically to benchmark generative tasks. The subset of {PixelProse} dataset, with approximately 30,000 samples, is used to train and test the hallucination evaluation model. Of these, 2,848 samples are used for testing the hallucination evaluation model’s performance, and the remaining samples are for training. We use LLaMA3 \cite{meta2023llama32} to generate negative image-caption pairs, including object hallucinations, as well as attribute and relation hallucinations. We apply our RL framework on two VLMs: LLaVA 7B \cite{liu2023llava} and InstructBLIP 7B \cite{huang2023visual} , with Vicuna backbone. The RL framework performs TTA on each AMBER test sample, leveraging feedback from the hallucination evaluation model. During RL-based inference, the learning rate is set to 2e-4 for InstructBLIP and 2e-3 for LLaVA. The beam size is set to 5, and the automatic adaptation process spans 5 steps per sample. After completing the TTA process for each test sample, the parameters of the VLM are reset to their initial state. This approach is crucial as it prevents the accumulation of sample-specific biases, maintains the generalizability of the pretrained model, and ensures that adaptations made for one sample do not negatively influence the performance on subsequent test samples.

The RL framework operates on a cluster of 4 A100 GPUs with 80 GB memory each, while the training of the hallucination evaluation model is performed on a single A100 GPU with 40 GB memory. The metrics are as follows \cite{wang2023llm}: 
 
\textbf{CHAIR} (Cumulative Hallucination Rate) measures the frequency of hallucinated objects in captions. It is defined as:
{\small\[
\text{CHAIR}(R) = 1 - \frac{\text{len}(R'_{\text{obj}} \cap A_{\text{obj}})}{\text{len}(R'_{\text{obj}})},
\]}
where \(R'_{\text{obj}}\) is the set of objects mentioned in the response, and \(A_{\text{obj}}\) represents the ground truth objects.

\textbf{Cover} quantifies the proportion of ground-truth objects covered in the caption. It is calculated as:
{\small\[
\text{Cover}(R) = \frac{\text{len}(R'_{\text{obj}} \cap A_{\text{obj}})}{\text{len}(A_{\text{obj}})}.
\]}

\textbf{Hal} (Hallucination Rate) indicates whether a response contains hallucinated objects. It is defined as:
{\small\[
\text{Hal}(R) =
\begin{cases} 
1 & \text{if } \text{CHAIR}(R) \neq 0, \\ 
0 & \text{otherwise}.
\end{cases}
\]}

\textbf{Cog} (Cognitive Overlap) evaluates the overlap between hallucinated objects in the responses and human cognitive tendencies. It is computed as:
{\small\[
\text{Cog}(R) = \frac{\text{len}(R'_{\text{obj}} \cap H_{\text{obj}})}{\text{len}(R'_{\text{obj}})},
\]}
where \(H_{\text{obj}}\) is the objects commonly hallucinated by humans.

\section{Results and Ablation Studies}
\label{sec:experiments_results}

As shown in Table \ref{tab:original_tta_comparison}, our framework demonstrates strong performance on both InstructBLIP and LLaVA 7B, with most metrics showing significant improvement after TTA. While LLaVA’s Cover metric drops slightly, likely due to the model narrowing the object range to improve accuracy and reduce hallucinations, this trade-off is acceptable given the overall gains in coherence and hallucination reduction. These results highlight the effectiveness of our approach in enhancing caption quality across diverse models.

\begin{table}[h!]
\caption{Comparison Across Models (w/ and w/o TTA)}
\label{tab:original_tta_comparison}
\renewcommand{\arraystretch}{1.5} 
\centering
\begin{tabular}{|c|c|c|c|c|c|}
\hline
\textbf{Model}         & \textbf{Type}  & \textbf{CHAIR (↓)} & \textbf{Cover (↑)} & \textbf{Hal (↓)} & \textbf{Cog (↓)} \\ \hline
\multirow{3}{*}{\shortstack{Instruct\\BLIP}} & w/o TTA  & 8.8                & 52.2               & 38.2             & 4.4              \\ \cline{2-6} 
                       & w/ TTA       & 6.5                & 54.1               & 31.6             & 3.3              \\ \cline{2-6}
                       & \textit{Change} & \textbf{-26.1\%}   & \textbf{+3.6\%}   & \textbf{-17.3\%}  & \textbf{-25.0\%} \\ \hline
\multirow{3}{*}{\shortstack{LLaVA\\7B }} & w/o TTA      & 7.8                & 51.0               & 36.4             & 4.2              \\ \cline{2-6} 
                       & w/ TTA       & 6.7                & 49.9               & 30.8             & 3.7              \\ \cline{2-6}
                       & \textit{Change} & \textbf{-14.1\%}   & \textbf{-2.2\%}   & \textbf{-15.4\%}  & \textbf{-11.9\%} \\ \hline
\end{tabular}
\end{table}

Table \ref{tab:model_parameters} highlights the efficiency and effectiveness of our framework. Both LLaVA 7B and InstructBLIP update only 0.0038\% and 0.0034\% of their total parameters, respectively, focusing exclusively on the layer normalization gamma parameters, and yet achieve significant performance gains. As shown in Table \ref{tab:original_tta_comparison}, these minimal updates yield substantial reductions in hallucination, including CHAIR, Hal, and Cog for both models, with only a slight drop on Cover for LLaVA 7B. The improvements demonstrate how even a few learnable parameters, strategically optimized, can result in large-scale enhancements in caption accuracy and reliability, further underscoring the lightweight yet powerful design of our framework.

\begin{table}[h!]
\caption{Learnable Parameters Across Models}
\label{tab:model_parameters}
\renewcommand{\arraystretch}{1.5} 
\centering
\begin{tabular}{|c|c|c|c|}
\hline
\textbf{Model}           & \textbf{Total Params} & \textbf{Learnable Params} & \textbf{Percent} \\ \hline
 CLIP$_{\text{Prompts + Triplet}}$               & 157M                  & 5.8M                      & 3.69                  \\ \hline
LLaVA 7B                    & 7.06B                 & 0.27M                     & 0.0038                \\ \hline
InstructBLIP             & 7.91B                 & 0.27M                     & 0.0034                \\ \hline
\end{tabular}
\end{table}

As shown in Table \ref{tab:vcd_vs_llava_tta}, the performance of LLaVA 7B with TTA demonstrates substantial improvements over the SOTA method Visual Contrastive Decoding (VCD) \cite{leng2024mitigating}  across key metrics. Specifically, CHAIR and Hal scores, indicating hallucination levels, are reduced by 91.7\% and 68.4\%, respectively, while the Cover score, reflecting object coverage, increases by 384.5\%. These significant gains demonstrate the capacity of TTA in mitigating hallucinations and improving object coverage, where \textbf{VCD struggles due to its limitations in addressing auto-regressive tasks that we have discussed in Section \ref{sec:related_work}}. Additionally, the lower Cog score achieved by TTA highlights its enhanced cognitive coherence, further establishing it as a superior method for hallucination mitigation.

\begin{table}[h!]
\caption{Comparison Between LLaVA 7B w/ VCD and LLaVA 7B w/ TTA}
\label{tab:vcd_vs_llava_tta}
\renewcommand{\arraystretch}{1.5} 
\centering
\begin{tabular}{|c|c|c|c|c|}
\hline
\textbf{Metric}        & \textbf{w/ VCD}   & \textbf{w/ TTA} & \textbf{Difference}  \\ \hline
CHAIR (↓)              & 80.5          & 6.7                   & \textbf{-91.7\%}    \\ \hline
Cover (↑)              & 10.3          & 49.9                  & \textbf{+384.5\%}   \\ \hline
Hal (↓)                & 97.5          & 30.8                  & \textbf{-68.4\%}    \\ \hline
Cog (↓)                & 9.2           & 3.7                   & \textbf{-59.8\%}    \\ \hline
\end{tabular}
\end{table}

Table \ref{tab:logit_comparison} highlights the impact of incorporating the NHP reward from the customized evaluation model. Both InstructBLIP and LLaVA 7B demonstrate consistent improvements across key metrics when using the NHP reward. CHAIR and Hal scores decrease, indicating reduced hallucination levels. Also, the decreased Cog scores reflects enhanced cognitive coherence. Although Cover scores slightly decrease for both models (-0.7\% for InstructBLIP and -0.4\% for LLaVA 7B), this trade-off is acceptable given the significant reduction in hallucinations. The decrease in Cover may result from the models narrowing their focus to align more strictly with factual content, thereby slightly limiting object coverage.

\begin{table}[h!]
\caption{Comparison between w/ logit and w/o logit Across InstructBLIP and LLaVA Models}
\label{tab:logit_comparison}
\renewcommand{\arraystretch}{1.5} 
\centering
\begin{tabular}{|c|c|c|c|c|c|}
\hline
\textbf{Model}         & \textbf{Logit}  & \textbf{CHAIR (↓)} & \textbf{Cover (↑)} & \textbf{Hal (↓)} & \textbf{Cog (↓)} \\ \hline
\multirow{3}{*}{\shortstack{Instruct\\BLIP}} & w/o  & 7.0                & 54.5               & 33.2             & 3.7              \\ \cline{2-6} 
                       & w/         & 6.5                & 54.1               & 31.6             & 3.3              \\ \cline{2-6}
                       & \textit{Change} & \textbf{-7.1\%}   & \textbf{-0.7\%}   & \textbf{-4.8\%}  & \textbf{-10.8\%} \\ \hline
\multirow{3}{*}{\shortstack{LLaVA\\7B}} & w/o        & 7.3                & 50.1               & 31.6             & 3.9              \\ \cline{2-6} 
                       & w/         & 6.7                & 49.9               & 30.8             & 3.7              \\ \cline{2-6}
                       & \textit{Change} & \textbf{-8.2\%}   & \textbf{-0.4\%}   & \textbf{-2.5\%}  & \textbf{-5.1\%} \\ \hline
\end{tabular}
\end{table}
As shown in Table \ref{tab:model_parameters}, the CLIP\textsubscript{Prompts + Triplet} model comprises only 5.8M learnable parameters, emphasizing its lightweight design. Before adopting it in our RL framework, we explored two variants. In CLIP\textsubscript{stacked}, CLS and EOS token features are combined and passed into stacked fully connected layers for classification. In contrast, CLIP\textsubscript{Prompts} introduces a learnable token that interacts with the frozen encoders, producing features then fed into stacked fully connected layers. Although both variants outperform the LLaVA models, neither surpasses our final CLIP\textsubscript{Prompts + Triplet} model, which builds upon CLIP\textsubscript{Prompts} by incorporating a triplet loss (see Section \ref{subsection:non_hallucination}) to further refine the representation space. As shown in Table \ref{tab:model_comparison}, CLIP\textsubscript{Prompts + Triplet} achieves the highest F1 scores for Object (86.5\%), Attribute (80.7\%), and Relation (75.5\%), surpassing the other models.  It highlights the synergistic effect of combining learnable prompts with triplet loss, enabling the CLIP-based model to deliver superior accuracy and robustness in hallucination evaluation while maintaining efficiency.

\begin{table}[h!]
\caption{Comparison of Models on Hallucination Detection (F1 Scores in \%)}
\label{tab:model_comparison}
\renewcommand{\arraystretch}{1.5} 
\centering
\begin{tabular}{|c|c|c|c|}
\hline
\textbf{Model}          & \textbf{Object (↑)} & \textbf{Attribute (↑)} & \textbf{Relation (↑)} \\ \hline
LLaVA 7B                & 15.0                & 26.6                   & 28.2                  \\ \hline
LLaVA 13B   \cite{liu2023llava}            & 10.2                & 41.3                   & 52.3                  \\ \hline
CLIP$_{\text{stacked}}$    & 84.9                & 77.8                   & 70.3                  \\ \hline
CLIP$_{\text{Prompts}}$     & 85.5                & 80.2                   & 71.8                  \\ \hline
CLIP$_{\text{Prompts + Triplet}}$  & \textbf{86.5}       & \textbf{80.7}          & \textbf{75.5}         \\ \hline
\end{tabular}
\end{table}

In contrast, the LLaVA models, including LLaVA 13B, show significantly lower scores in hallucination detection, particularly for object hallucinations (15.0\% for LLaVA 7B and 10.2\% for LLaVA 13B), highlighting the inherent difficulty of this task. We then retrained and tested the LLaVA 7B model on the same dataset, achieving results closer to our CLIP$_{\text{Prompts + Triplet}}$ model. However, its large parameter count incurs high computational costs, and its inability to generate CLIP scores for reward calculations in our RL framework makes it less suitable for our approach. Therefore, we focused on the lightweight and efficient CLIP-based approach, which better aligns with the requirements of our framework.

Additionally, as the hallucination evaluation model was trained on three types of hallucinations: object, attribute, and relation, it has the potential to be extended for broader hallucination detection tasks beyond the current scope. This versatility positions it as a promising foundation for future research and applications focused on comprehensive hallucination mitigation across diverse scenarios.

\section{Conclusion and Future Directions}
\label{sec:conclusion}

This work tackles the critical challenge of mitigating hallucinations in VLMs by introducing a novel TTA framework using RL. By updating only the lightweight layer normalization gamma parameters and incorporating a customized CLIP-based hallucination evaluation model, our approach effectively reduces object hallucinations during inference. Experiments demonstrate substantial performance improvements across multiple metrics on SOTA VLMs, showcasing the framework’s robustness and efficiency. In future work, we aim to extend this framework to attribute and relationship hallucinations, leveraging the capabilities of the customized CLIP model for a more comprehensive solution. Additionally, we plan to explore its application in discriminative tasks and adversarial training setups, where the VLM acts as a generative model and the evaluation model serves as a discriminator. These advancements may further enhance hallucination mitigation in VLMs, broadening their real-world applicability.

\bibliographystyle{IEEEtran}


\bibliography{IEEEabrv,icme2023_ref_file}

\end{document}